\documentclass{article}
\usepackage{graphicx}

\title{Saddle-splay modulus \\ of a particle-laden fluid interface}

\author{S. V. Lishchuk}

\date{
Materials and Engineering Research Institute,
Sheffield Hallam University,
Sheffield S1~1WB, United Kingdom
}

\begin{document}

\maketitle

\begin{abstract}
The scaled-particle theory equation of state for the two-dimensional hard-disk fluid on a curved
surface is proposed and used to determine the saddle-splay modulus of a particle-laden fluid
interface. The resulting contribution to saddle-splay modulus, which is caused by thermal motion of
the adsorbed particles, is comparable in magnitude with the saddle-splay modulus of a simple fluid
interface.
\end{abstract}

%%%%%%%%%%%%%%%%%%%%%%%%%%%%%%%%%%%%%%%%%%%%%%%%%%%%%%%%%%%%%%%%%%%%%%%%%%%%%%%%%%%%%%%%%%%%%%%%%%%%

\section{Introduction}

The surface free energy density of fluid interfaces depends upon their curvature. This dependence
affects the nucleation in liquids \cite{Tolman:1949-333,Schmelzer:1996-657,Baidakov:1999-469}, and
has important role in determining the structure and dynamics of the systems with complex fluid
interfaces, such as membranes or surfactants \cite{Safran:STSIM}.

For small curvature of the interface, the dependence of the surface free energy $f$ upon the geometry
of the interface is conveniently described by the Helfrich curvature expansion
\cite{Helfrich:1973-693}
\begin{equation}
\label{eq:Helfrich}
f=\sigma+2\kappa(H-H_0)^2+\bar\kappa K
\end{equation}
In this equation the geometry of the interface is characterized by mean curvature
$H=\frac12(1/R_1+1/R_2)$ and Gaussian curvature $K=1/(R_1R_2)$, $R_1$ and $R_2$ being principal radii
of curvature of the interface. The surface tension $\sigma$, bending modulus $\kappa$, spontaneous
curvature $H_0$, and saddle-splay modulus (or Gaussian rigidity) $\bar\kappa$ are the material
parameters of the interface. By virtue of Gauss-Bonnet theorem, the contribution of the last term in
Eq.~(\ref{eq:Helfrich}) into the total free energy of the system depends on the topology of the
system. Indeed, the value of the saddle-splay modulus affects the processes which involve changes in
the topology of fluid interfaces
\cite{Safran:1991-2903,Gompper:1998-2284,Le:2000-379,Jung:2002-15318,Siegel:2004-366,Kozlovsky:2004-2508}.

An interesting example of a system which can be macroscopically viewed as a complex fluid interface
is the fluid interface laden with colloidal micro- or nanoparticles.
To minimize total interfacial energy, particles suspended in a bulk fluid self-assemble on the
fluid interface \cite{Pieranski:1908-569}. This process, first observed by Ramsden in 1903
\cite{Ramsden:1903-156}, has recently attracted significant scientific attention
\cite{Binks:CPLI,Binks:2002-21,Boker:2007-1231,Menner:2007-2398}. It has also potential for a range
of novel applications
\cite{Dinsmore:2002-1006,Strohm:2004-2667,Neirinck:2007-57,Neirinck:2008-246,Torres:2008-123}.

On the scale large compared to the size of the adsorbed particles, a particle-laden interface may
be viewed as continuous. If the interface is isotropic on this scale, the interfacial free energy
can be described by Eq.~(\ref{eq:Helfrich}), and the interface can be characterized by the material
parameters $\sigma$, $\kappa$, $H_0$, and $\bar\kappa$.

The present letter is devoted to the study of the saddle-splay modulus $\bar\kappa$ of a
particle-laden fluid interface at low surface concentration of the adsorbed particles. In this case
we can represent the interface as a two-dimensional fluid on a curved surface. The main contribution
to the interaction between particles at low concentration comes from the excluded volume (different
particles cannot occupy the same space). Hence we approximate the system by a two-dimensional
hard-disk fluid on a curved surface.

Hard-disk fluids in curved geometry were used before to study packing of disks
\cite{Schreiner:1982-379,Nelson:1983-982,Rubinstein:1983-6377,Modes:2008-041125}, ordering phase
transition \cite{Giarritta:1992-456}, topological defects \cite{Giarritta:1993-649}, and as a model
of glass-forming liquids \cite{Nelson:1983-982,Rubinstein:1983-6377,Sausset:2008-155701}. Several
equations of state were proposed for hard-disk fluids in spherical
\cite{Tobochnik:1988-5824,Lishchuk:2006-266} and hyperbolic
\cite{Modes:2007-235701,Modes:2008-041125,DeHaro:2008-116101} geometries.

In the present work we shall use scaled-particle theory (SPT) \cite{Reiss:1959-369} to derive the
equation of state of two-dimensional hard-disk fluid on a curved surface. We shall then use the
resulting equation of state to determine saddle-splay modulus $\bar\kappa$ for particle-laden fluid
interface at low concentration of the adsorbed particles.

%%%%%%%%%%%%%%%%%%%%%%%%%%%%%%%%%%%%%%%%%%%%%%%%%%%%%%%%%%%%%%%%%%%%%%%%%%%%%%%%%%%%%%%%%%%%%%%%%%%%

\section{Saddle-splay modulus}

In accordance with Eq.~(\ref{eq:Helfrich}), the saddle-splay modulus is given by the derivative of
the surface free energy density with respect to Gaussian curvature,
\begin{equation}
\label{eq:kappa-def0}
\bar\kappa=\left.\frac{\partial f}{\partial K}\right|_{K=0}.
\end{equation}
Using the expression for the excess free energy
\begin{equation}
\frac{\beta F^{\mathrm{ex}}}N=\int_0^\rho\frac{Z-1}\rho d\rho,
\end{equation}
where
\begin{equation}
Z\equiv\frac{\beta P}\rho
\end{equation}
is the compressibility factor, $\rho=N/A$ is the number density (number of particles per unit area),
$P$ is pressure, $\beta=1/k_BT$ is the inverse temperature, we may represent
Eq.~(\ref{eq:kappa-def0}) in the form
\begin{equation}
\label{eq:kappa-def1}
\bar\kappa=\frac\rho\beta\int_0^\rho\frac1\rho\left(\frac{\partial Z}{\partial K}\right)_{K=0}d\rho,
\end{equation}
where the derivative is taken at constant particle density $\rho$.

Equation (\ref{eq:kappa-def1}) can be used to calculate saddle-splay modulus of the interface from
the curvature dependence of the compressibility factor, which is generally given by the equation of
state of the system. We shall use the SPT equation of state for a hard-disk fluid on a curved
surface, which is derived in the following sections.

%%%%%%%%%%%%%%%%%%%%%%%%%%%%%%%%%%%%%%%%%%%%%%%%%%%%%%%%%%%%%%%%%%%%%%%%%%%%%%%%%%%%%%%%%%%%%%%%%%%%

\section{SPT equation of state for hard disks}

Scaled-particle theory was originally
developed by Reiss {\em et al} \cite{Reiss:1959-369} and further improved afterwards
\cite{Gibbons:1969-81,TullySmith:1970-4015,Mandell:1975-113,Heying:2004-19756,Siderius:2007-144502}.
Applied to the case of hard disks on a 2D plane, SPT leads to a particularly simple equation of
state which is nevertheless in good agreement with computer simulation results throughout most
of the fluid range of densities \cite{Helfand:1961-1037,Cotter:1972-3356}.

SPT for two-dimensional hard-disk fluids in its simplest form can be summarized as follows
(see textbook~\cite{Hansen:TSL} for more details). The reversible work $W(R_0)$ is considered which
is required to create a circular cavity of radius $R_0$ in the fluid of hard disks of radius $R$.
The assumption is made that for $R_0>0$, $W(R_0)$ is given by a polynomial in $R_0$
\begin{equation}
\label{eq:large-cavities}
W(R_0)=w_0+w_1R_0+S(R_0)P,\quad R_0\ge0.
\end{equation}
The last term $S(R_0)P$ ($S(R)$ being the area of the disk of radius $R$), which is dominant for
large cavities ($R_0\gg R$), follows from thermodynamics. For small cavities ($0\le R+R_0\le R$),
$W(R_0)$ can be written in form
\begin{equation}
\label{eq:small-cavities}
W(R_0)=-k_BT\ln\left[1-\rho S(R_0+R)\right],\; -R\le R_0\le0.
\end{equation}
The coefficients $w_0$ and $w_1$ are then determined by requiring the work $W(R_0)$ and its
derivative $W'(R_0)$, given by Eqs~(\ref{eq:large-cavities}) and (\ref{eq:small-cavities}), to be
continuous at $R_0=0$. The explicit expression for the excess chemical potential of the fluid,
$\mu^\mathrm{ex}=W(R)$, can be determined from Eq.~(\ref{eq:large-cavities}), and subsequently used
to write the SPT equation of state.

In the case of the flat surface, the area of the disk is
\begin{equation}
\label{eq:S-flat}
S(R)=\pi R^2.
\end{equation}
The corresponding
values of the coefficients $w_i$ are given by
\begin{equation}
\beta w_0=-\ln(1-\eta),
\quad
\beta w_1=\frac{2\pi\rho R}{1-\eta},
\end{equation}
where $\eta=\pi R^2\rho$ is the hard-disk packing fraction. The chemical potential of the fluid,
$\mu$, is given by
\begin{equation}
\label{eq:mu-flat}
\beta\mu=\ln\Lambda^2\rho-\ln(1-\eta)+\frac{2\eta}{1-\eta}+\frac{\beta P\eta}\rho,
\end{equation}
where $\Lambda$ is de Broglie thermal wavelength. The SPT equation of state is then obtained from
Eq.~(\ref{eq:mu-flat}) and the thermodynamic relation
\begin{equation}
\label{eq:thermodynamic}
\frac{\partial P}{\partial\rho}=\rho\frac{\partial\mu}{\partial\rho},
\end{equation}
and has the form reported by Helfand {\em et al} \cite{Helfand:1961-1037}:
\begin{equation}
\label{eq:EOS-flat}
Z=\frac1{(1-\eta)^2}.
\end{equation}

%%%%%%%%%%%%%%%%%%%%%%%%%%%%%%%%%%%%%%%%%%%%%%%%%%%%%%%%%%%%%%%%%%%%%%%%%%%%%%%%%%%%%%%%%%%%%%%%%%%%

\section{SPT equation of state for hard disks on a curved surface}

SPT equation of state for a hard-disk fluid on a curved surface can be obtained in the same way as
in the flat case described above. The difference is that expression (\ref{eq:S-flat}) for the area
of the disk of radius $R$ is no longer valid on a curved surface. For small Gaussian curvature
($K\ll1/R^2$) we shall replace it by the formula for the area of a geodesic disk on two-dimensional
Riemannian manifold, obtained by Bertrand and Diguet in 1848 \cite{Bertrand:1848-80},
\begin{equation}
\label{eq:S-curved}
S(R)=\pi R^2(1-\xi)+o(\xi),
\end{equation}
where we have introduced the dimensionless quantity
\begin{equation}
\xi\equiv\frac{KR^2}{12}.
\end{equation}
Requiring work $W(R_0)$ and its derivative $W'(R_0)$, as given by Eqs~(\ref{eq:large-cavities}) and
(\ref{eq:small-cavities}), to be continuous at $R_0=0$, we obtain the following expressions for the
coefficients $w_i$,
\begin{equation}
\beta w_0=-\ln[1-\pi R^2\rho(1-\xi)],
\end{equation}
\begin{equation}
\beta w_1=\frac{2\pi R\rho(1-2\xi)}{1-\pi R^2\rho(1-\xi)},
\end{equation}
and the chemical potential,
\begin{eqnarray}
\nonumber
\beta\mu&=&\ln\Lambda^2\rho-\ln[1-\eta(1-\xi)]
\\
\label{eq:mu-curved}
&+&\frac{2\eta(1-2\xi)}{1-\eta(1-\xi)}+\frac{\beta P\eta(1-\xi)}\rho,
\end{eqnarray}
Equations (\ref{eq:mu-curved}) and (\ref{eq:thermodynamic}) lead to the following form of the SPT
equation of state for hard-disk fluid on a curved surface:
\begin{equation}
\label{eq:EOS-curved}
Z=\frac{1-\eta\xi}{[1-\eta(1-\xi)]^2}.
\end{equation}

\begin{figure}[h]
\begin{center}
\includegraphics[width=\columnwidth,keepaspectratio]{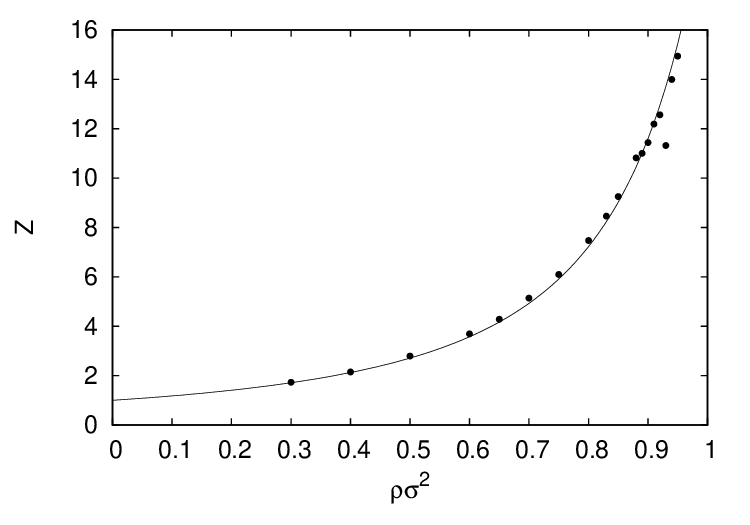}
\end{center}
\caption{
Compressibility factor $Z$ as a function of reduced number density of the fluid $\rho\sigma^2$,
where $\sigma\equiv2R$ is particle diameter. Circles represent Monte Carlo results for $N=400$ hard
disks on a sphere \cite{Giarritta:1992-456}, line corresponds to SPT equation of state
(\ref{eq:EOS-curved}).
}
\label{fig:giarritta92}
\end{figure}

Figure~\ref{fig:giarritta92} demonstrates satisfactory agreement of the compressibility factor $Z$
calculated from Eq.~(\ref{eq:EOS-curved}) with the Monte Carlo results for hard disks on a sphere
reported by Giarritta {\em et al} \cite{Giarritta:1992-456}. In the case of zero Gaussian curvature
($\xi=0$) Eq.~(\ref{eq:EOS-curved}) coincides with Eq.~(\ref{eq:EOS-flat}).

%%%%%%%%%%%%%%%%%%%%%%%%%%%%%%%%%%%%%%%%%%%%%%%%%%%%%%%%%%%%%%%%%%%%%%%%%%%%%%%%%%%%%%%%%%%%%%%%%%%%

\section{Saddle-splay modulus from SPT equation of state}

The expression for saddle-splay modulus is obtained by substituting the compressibility factor given
by the equation of state, Eq.~(\ref{eq:EOS-curved}), into formula (\ref{eq:kappa-def1}). The result
is
\begin{equation}
\label{eq:kappa}
\bar\kappa_{SPT}=-k_BT\frac{\eta^2(3-2\eta)}{12\pi(1-\eta)^2}.
\end{equation}
Note that although using the truncated series in $R$ given by the formula (\ref{eq:S-curved}) for
the area of the large disk in the expression (\ref{eq:large-cavities}) is generally not justified,
it is still suitable for our purpose of calculating saddle-splay modulus since we are interested in
the limit $K\rightarrow0$.

\begin{figure}[h]
\begin{center}
\includegraphics[width=\columnwidth,keepaspectratio]{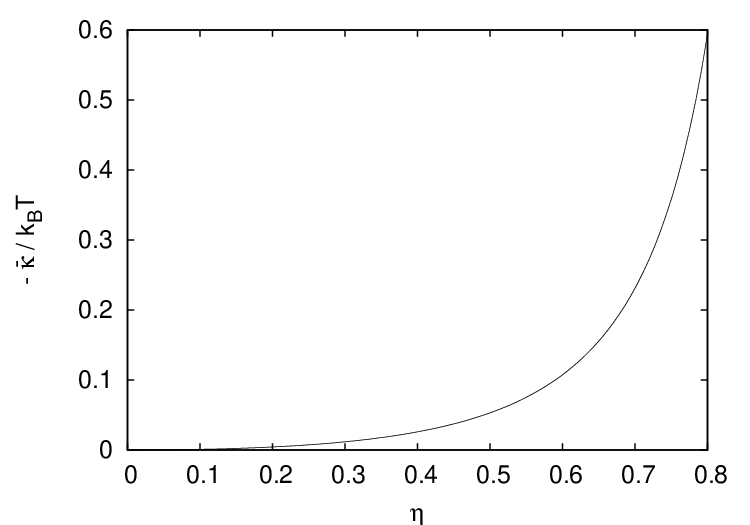}
\end{center}
\caption{
The dependence of the saddle-splay modulus $\bar\kappa$ upon disk packing fraction $\eta$,
calculated using Eq.~(\ref{eq:kappa}).
}
\label{fig:kappa}
\end{figure}

The dependence of the saddle-splay modulus $\bar\kappa $ upon the disk packing fraction $\eta$,
given by Eq.~(\ref{eq:kappa}), is presented in Figure~\ref{fig:kappa}. The value of the saddle-splay
modulus for particle-laden interfaces appears to be smaller than the values $|\bar\kappa|\sim10k_BT$
typical for lipid monolayers \cite{Marsh:2006-146}), but is comparable to the value
$|\bar\kappa|\approx\frac12k_BT$ for the surfaces of simple fluids \cite{VanGiessen:2002-302}.

%%%%%%%%%%%%%%%%%%%%%%%%%%%%%%%%%%%%%%%%%%%%%%%%%%%%%%%%%%%%%%%%%%%%%%%%%%%%%%%%%%%%%%%%%%%%%%%%%%%%

\section{Conclusion}

The main message of this letter is that the thermal motion of the particles adsorbed on a fluid
interface contributes to the saddle-splay modulus of the interface. This result may have
implications in the structure and dynamics particle-laden systems that allow topological changes,
for example, fusion of particles in Pickering emulsions, or structural reorganization in
particle-stabilized foams.

The simplest version of the scaled-particle theory allows construction of a rather simple equation
of state for a hard-disk fluid on a curved surface. In order to improve the formula obtained for
the saddle-splay modulus of a particle-laden fluid interface, it seems reasonable to attempt to
construct the equation of state that gives more accurate dependence of the compressibility factor
with respect to Gaussian curvature of the interface, which can be verified by using the virial
expansion on the curved surface or the computer modelling of the system.

The result can also be extended by taking into account the influence on the value of saddle-splay
modulus of other contributions to the interparticle interaction, such as capillary, electrostatic,
van der Waals {\em etc}, as well as the role of particles' anisotropy. The prediction of the elastic
properties of the interfaces with large concentration of particles, in which two-dimensional solid
structure forms, presents another interesting and more complicated problem.

%%%%%%%%%%%%%%%%%%%%%%%%%%%%%%%%%%%%%%%%%%%%%%%%%%%%%%%%%%%%%%%%%%%%%%%%%%%%%%%%%%%%%%%%%%%%%%%%%%%%

\section*{Acknowledgement}

I thank Prof.\ C.M.~Care for discussion of the results.

%%%%%%%%%%%%%%%%%%%%%%%%%%%%%%%%%%%%%%%%%%%%%%%%%%%%%%%%%%%%%%%%%%%%%%%%%%%%%%%%%%%%%%%%%%%%%%%%%%%%

%\bibliographystyle{unsrt}
%\bibliography{spt}

%%%%%%%%%%%%%%%%%%%%%%%%%%%%%%%%%%%%%%%%%%%%%%%%%%%%%%%%%%%%%%%%%%%%%%%%%%%%%%%%%%%%%%%%%%%%%%%%%%%%

\end{document}